\documentclass[twocolumn,preprintnumbers,amsmath,aps]{revtex4}
\usepackage{graphicx}
\usepackage{dcolumn}
\usepackage{bm}
\usepackage{subfigure}
\usepackage[usenames]{color}
\begin{document}
%%%%%%%%%%%%%%%%%%%%%%%%%%%%
\newcommand{\kvec}{\mbox{{\scriptsize {\bf k}}}}
%%%%%%%%%%%%%%%%%%%%%%%%%%%%
\def\eq#1{(\ref{#1})}
\def\fig#1{\ref{#1}}
\def\tab#1{\ref{#1}}
%%%%%%%%%%%%%%%%%%%%%%%%%%%%
\title{Superconducting properties of lithium-decorated bilayer graphene}
\author{D. Szcz{\c{e}}{\'s}niak}\email{dszczesniak@qf.org.qa}
%%%%%%%%%%%%
\affiliation{Qatar Environment and Energy Research Institute, Qatar Foundation, PO Box 5825, Doha, Qatar}
%%%%%%%%%%%%
%%%%%%%%%%%%
\date{\today} 
\begin{abstract}
%%%%%%%%%%%%%%%%%%%%%%%%%%%%%%%%%%%%%%%%%%%%%%%%%%%

Present study provides a comprehensive theoretical analysis of the superconducting phase in selected lithium-decorated bilayer graphene nanostructures. The numerical calculations, conducted within the Eliashberg formalism, give quantitative estimations of the most important thermodynamic properties such as the critical temperature, specific heat, critical field and others. It is shown that discussed lithium-graphene systems present enhancement of their thermodynamic properties comparing to the monolayer case {\it e.g.} the critical temperature can be raised to $\sim 15$ K. Furthermore, estimated characteristic thermodynamic ratios exceed predictions of the Bardeen-Cooper-Schrieffer theory suggesting that considered lithium-graphene systems can be properly analyzed only within the strong-coupling regime.

%%%%%%%%%%%%%%%%%%%%%%%%%%%%%%%%%%%%%%%%%%%%%%%%%%%
\end{abstract}
\maketitle

%%%%%%%%%%%%%%%%%%%%%%%%%%%%%%%%%%%%%%%%%%%%%%%%%%%
\section{Introduction}
%%%%%%%%%%%%%%%%%%%%%%%%%%%%%%%%%%%%%%%%%%%%%%%%%%%

Among all carbon allotropes \cite{yang1}, \cite{szczesniak1}, \cite{noorden}, \cite{hirsch}, graphene, a single atomic carbon layer \cite{novoselov}, is of special interest for a wide scientific community \cite{castro1}. The outstanding electronic \cite{castro2}, optical \cite{bonaccorso}, thermal \cite{balandin}, and mechanical \cite{lee} properties of pristine graphene already established this material as a promising candidate for a versatile variety of future applications in nanotechnology \cite{ferrari}.

At present, one of the most popular application domains concerns the possibility of using graphene in nanoelectronics \cite{fiori}. Even by exploiting just the fundamental electronic properties of this material such as the high carrier mobility, the perfect charge-carrier confinement as well as the highly efficient carrier-carrier scattering, it is possible to use graphene in future ambipolar transistors \cite{yang2}, graphene/silicon hybrid systems \cite{huang}, ultrafast photodetectors \cite{xia}, as well as the highly-efficient solar cells \cite{tielrooij}.

All these inherent electronic features of unmodified graphene stems from its intrinsically gapless characteristic, protected by the inversion and time-reversal symmetry. However for some applications such low-energy dynamics of electrons constitute at the same time one of the crucial drawbacks of this material. In particular, along with the low density of states at the Fermi level, highly energetic in-plane vibrations, and lack of the coupling between the in-plane electronic states and the out-of-plane vibrations, they prevent the manifestation of the quantum phenomenon of superconductivity. If possible, novel graphene-based superconductors may lead to the new generation of the superconductor-quantum dot devices \cite{fan} or low-dimensional superconducting transistors \cite{franceschi}.

Until now the possibility of inducing superconductivity in graphene was addressed mainly by the theoretical studies. These investigations initially were considering unconventional pairing mechanisms around the characteristic Dirac points at the Fermi energy \cite{kopnin}, \cite{uchoa} or in the range of higher energies when the Fermi energy is raised to the vicinity of the van Hove singularity \cite{nandkishore}. On the contrary, an early experimental investigations were concentrating on the proximity effect driven supercurrents \cite{heersche}.

Later on it was predicted that the conventional electron-phonon interactions may also lead to the induction of the superconducting state in graphene \cite{einenkel}. In this spirit, the biggest attention in recent years was devoted to strengthening the electron-phonon interactions by chemical doping. The first interesting attempt was given in \cite{savini}, suggesting a high-temperature superconductivity in graphane (a strongly hydrogenated graphene) under hole doping.

Another theoretical proposals were devoted to doping graphene with lithium atoms. The advent of this direction of research is marked by the recent work of Profeta {\it et al.} \cite{profeta} where authors discussed possibility of inducing superconducting phase in graphene due to the removal of the quantum confinement and hence generation of the intralayer lithium electronic state at the Fermi level. In what follows, the desirable rise of the density of states at the Fermi level has been achieved, resulting in the strong-coupling character of the superconducting phase in such monolayer lithium-decorated graphene structures \cite{szczesniak2}. 

Further theoretical investigations were considering enhancement of the superconducting properties in graphene-lithium materials due to the influence of the hexagonal boron nitride substrate \cite{kaloni} or applied strain \cite{pesic}, as well as engineering the few-layer structures \cite{guzman}. This is a general trend in the domain of superconductivity which aims to achieve the highest possible value of critical temperature ($T_{C}$) in the given material (please see for example \cite{szczesniak4}, \cite{durajski1}, \cite{szczesniak6}). In particular, proposition given in \cite{guzman} should be of crucial interest for the scientific community, since bilayer lithium-decorated graphene structures are already experimentally proved to be thermodynamically stable \cite{sugawara}.

In this context, present paper provides systematic analysis of the thermodynamics of the superconducting phase in selected graphene-lithium nanostructures. In particular, calculations are conducted for the most optimal lithium-covered bilayer graphene structures which are predicted to present the enhancement of the superconducting properties comparing to the monolayer case discussed in \cite{profeta}, \cite{szczesniak2}. Due to the fact that considered structures are characterized by the relatively high electron-phonon coupling constant ($\lambda$) values the calculations presented in this work are carried out within the Eliashberg formalism \cite{eliashberg}, a strong-coupling generalization of the classical Bardeen-Cooper-Schrieffer theory \cite{bardeen1}, \cite{bardeen2}.

%%%%%%%%%%%%%%%%%%%%%%%%%%%%%%%%%%%%%%%%%%%%%%%%%%%
\section{Theoretical model}
%%%%%%%%%%%%%%%%%%%%%%%%%%%%%%%%%%%%%%%%%%%%%%%%%%%

\begin{figure}[h]
\includegraphics[width=\columnwidth]{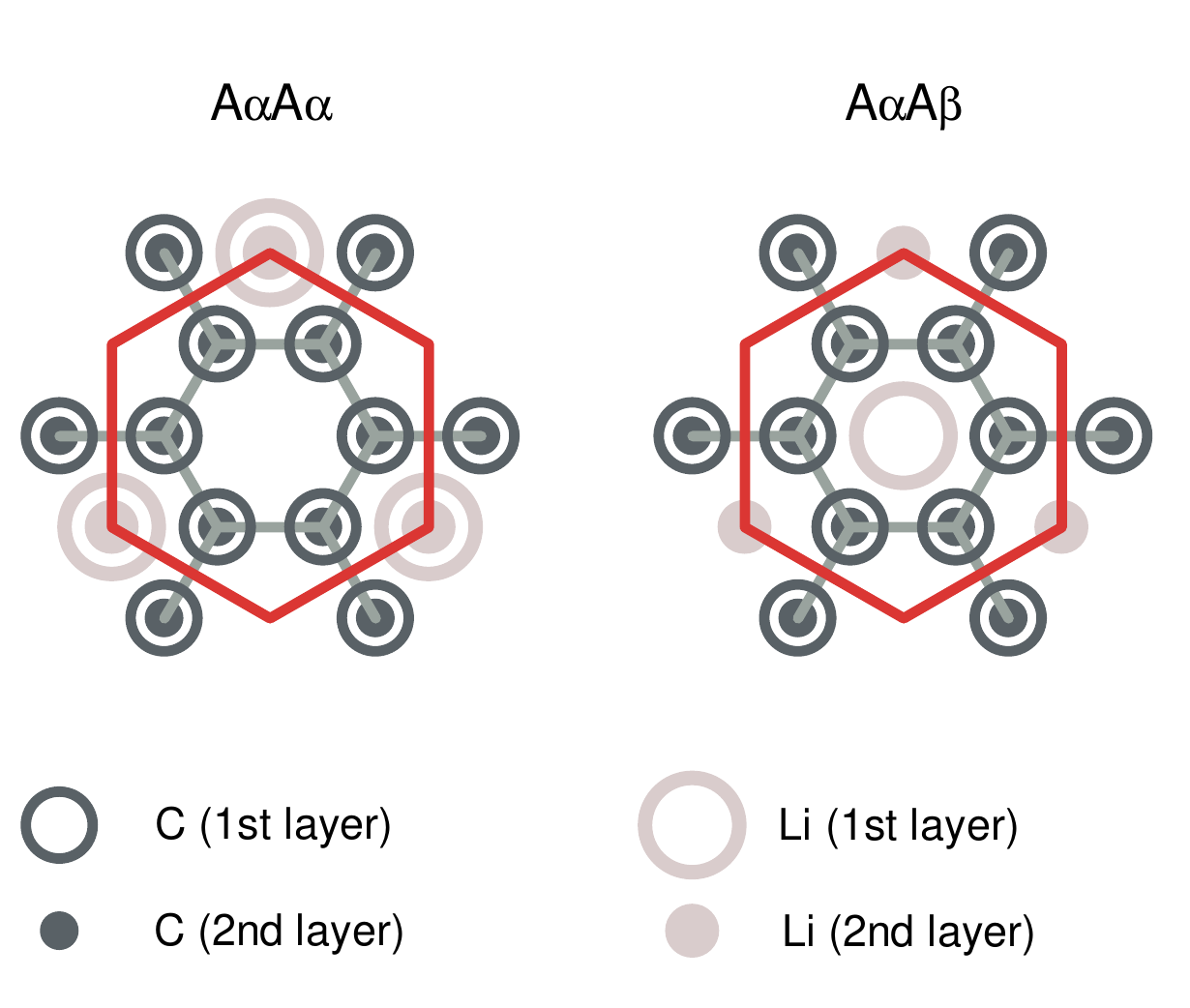}
\caption{Stacking schemes for $A \alpha A \alpha$ and $A \alpha A \beta$ lithium decorated bilayer graphene. The Wigner-Seitz primitive unit cells are marked in red.}
\label{fig1}
\end{figure}

The primitive unit cells of the two lithium-decorated bilayer graphene structures, considered in the present paper, are depicted in Fig \ref{fig1}. In general, analyzed systems consist of two graphene and two lithium layers which are ordered in an alternating manner. The two graphene layers are always arranged, with respect to each other, so that the A and B sub-lattice carbon atoms of one layer are exactly above the corresponding A and B atoms of the second layer (the so-called $AA$-stacking sequence). 

The difference between both systems is viewed in the terms of the stacking schemes of the two lithium layers. In the first case, lithium atoms in both layers occupy the same hollow sites above and below the second graphene layer (stacking sequence $A \alpha A \alpha$). In the latter case lithium layers are shifted to each other by the lattice vector of the graphene layer (stacking sequence $A \alpha A \beta$).

As already mentioned, both systems are theoretically predicted to be the phonon-mediated superconductors with the electron-phonon coupling constant values exceeding the BCS limit ($\alpha < 0.3$) \cite{bardeen2}. In particular, $\lambda=0.65$ and $\lambda=0.86$ for $A \alpha A \alpha$  and  $A \alpha A \beta$ cases, respectively. In what follows, the thermodynamic properties of these systems are described, throughout the present work, in the framework of the isotropic Eliashberg equations \cite{eliashberg}. The isotropic character of the Elishberg equations is caused by the single-band character of the electron-phonon spectral functions ($\alpha^2F(\Omega)$) presented in \cite{guzman}, which is adopted in the present analysis.

In order to solve the set of the isotropic Eliashberg equations the iterative method presented in \cite{szczesniak2}, \cite{durajski2}, \cite{szczesniak9} is used. The calculations are conducted on the imaginary axis and in the mixed representation, taking into consideration the 1100 Matsubara frequencies: $\omega_{m}\equiv\frac{\pi}{\beta}(2m-1)$, where $\beta\equiv 1/k_{B}T$, with $k_{B}$ denoting the Boltzmann constant. This numerical precision allows to ensure the stability of the solutions for $T\geq T_{0}\equiv 1.75$ K.

To this end, the electron departing correlations are modeled by the Coulomb pseudopotential of the value of 0.1, and the cutoff frequency for the calculations is assumed to be $\omega_{c}=10\Omega_{\rm max}$, where $\Omega_{\rm max}$ denotes the maximum phonon frequency equals to $193.25$ meV and to $181.68$ meV for $A \alpha A \alpha$ and $A \alpha A \beta$ cases, respectively.

%%%%%%%%%%%%%%%%%%%%%%%%%%%%%%%%%%%%%%%%%%%%%%%%%%%
\section{Results and discussion}
%%%%%%%%%%%%%%%%%%%%%%%%%%%%%%%%%%%%%%%%%%%%%%%%%%%

%%%%%%%%%%%%%%%%%%%%%%%%%%%%%%%%%%%%%%%%%%%%%%%%%%%
\subsection{Order parameter and superconducting transition temperature}
%%%%%%%%%%%%%%%%%%%%%%%%%%%%%%%%%%%%%%%%%%%%%%%%%%%

\begin{figure}[ht]
\includegraphics[width=\columnwidth]{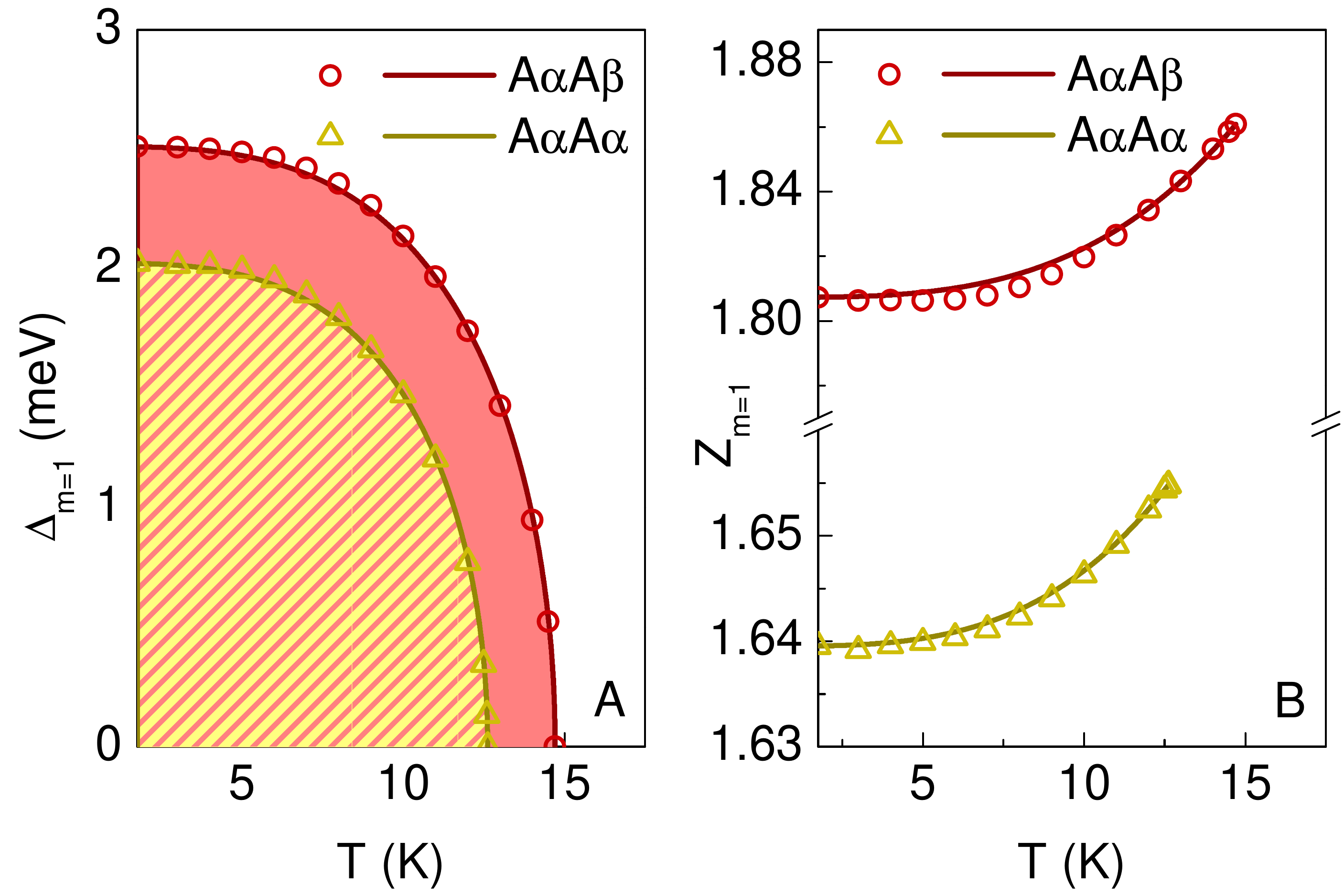}
\caption{The dependance of the maximum value of the order parameter ($\Delta_{m=1}$) (A) and the wave function renormalization factor ($Z_{m=1}$) (B) on temperature, for the $A \alpha A \alpha$ and $A \alpha A \beta$ stacking cases. The open symbols mark the exact Eliashberg results whereas solid lines denote results obtained with the fitting formulas \ref{eq1} and \ref{eq2} in the (A) and (B) subfigures, respectively. The shaded areas in (A) give the superconducting phase existence regions for the $A \alpha A \alpha$ and $A \alpha A \beta$ configurations.}
\label{fig2}
\end{figure}

The superconducting transition temperature can be determined quantitatively on the basis of the behavior of the maximum value of the order parameter ($\Delta_{m=1}$) by solving the Eliashberg equations on the imaginary axis. In details, $T_{C}$ is equal to the temperature at which $\Delta_{m=1}$ reaches the value of zero ($\Delta_{m=1}(T_{C})=0$). In Fig \ref{fig2} (A), the dependence of the maximum value of the order parameter on temperature is presented for two considered stacking cases of the lithium-decorated bilayer graphene. In this figure, open symbols represent the exact Eliashberg solutions, whereas solid lines stand for the results obtained by using the following fitting formula:
\begin{equation}
\label{eq1}
\Delta_{m=1}=\Delta_{m=1}\left(0\right)\sqrt{1-\left(\frac{T}{T_{C}}\right)^{\eta}}, 
\end{equation}
where $\Delta_{m=1}\left(0\right)\equiv\Delta_{m=1}\left(T_{0}\right)$ is equal to 2.02 meV for the $A \alpha A \alpha$ stacking and 2.51 meV for the $A \alpha A \beta$ stacking, whereas $\eta=3.25$ is the fitting paramteter.

In this context, the estimated critical temperature values are equal to 12.62 K and 14.71 K for the $A \alpha A \alpha$ and $A \alpha A \beta$ systems, respectively. In what follows the increase in the critical temperature value is observed when comparing to the monolayer LiC$_6$ case (denoted here as a $A \alpha$ stacking). This fact can be quantitatively summarized by the following ratios: $T_{C}^{A \alpha A \alpha}/{T_{C}^{A \alpha}}=1.48$ and $T_{C}^{A \alpha A \beta}/{T_{C}^{A \alpha}}=1.72$, where $T_{C}^{A \alpha}=8.55$ K \cite{szczesniak2}.

Results presented in Fig \ref{fig2} (A) can be next used to determine, in the first approximation, the value of the energy band gap at the Fermi level ($2\Delta_{m=1}(0)$). In the case of the $A \alpha A \alpha$ and $A \alpha A \beta$ type systems the $2\Delta_{m=1}(0)$ is equal to 4.05 meV and 5.02 meV, respectively.

Furthermore, by analyzing one of the $\Delta_{m=1}$ function components, namely the maximum value of the wave function renormalization factor ($Z_{m=1}$), another approximate estimation of the physical observable can be given. In particular, the $Z_{m=1}$ as a function of temperature (as presented in Fig \ref{fig2} (B)) describes the electron effective mass ($m^{\star}_{e}$) dependence on the temperature. The corresponding relation is written as $m^{\star}_{e}\simeq Z_{m=1} m_{e}$, where $m_{e}$ denotes the band electron mass. In Fig \ref{fig2} (B), similarly as in the case of the $\Delta_{m=1}$ function, the open symbols correspond to the exact Eliashberg results and the solid lines are plotted with the help of the following formula:
\begin{eqnarray}
\label{eq2}\nonumber
Z_{m=1}&=&\left[Z_{m=1}\left(T_{C}\right) - Z_{m=1}\left(T_{0}\right) \right] \left(\frac{T}{T_{C}}\right)^{\eta}\\
&+& Z_{m=1}\left(T_{0}\right), 
\end{eqnarray}
where $Z_{m=1}\left(T_{C}\right)$ amounts 1.65 for the $A \alpha A \alpha$ case and 1.86 for the $A \alpha A \beta$  case. Moreover, the $Z_{m=1}\left(T_{0}\right)$ is equal to 1.64 and 1.86 for the $A \alpha A \alpha$ and $A \alpha A \beta$ systems, respectively.

%%%%%%%%%%%%%%%%%%%%%%%%%%%%%%%%%%%%%%%%%%%%%%%%%%%%%%%
\subsection{Physical value of the energy gap at the Fermi level and the electron effective mass}
%%%%%%%%%%%%%%%%%%%%%%%%%%%%%%%%%%%%%%%%%%%%%%%%%%%%%%%

In order to determine precisely the physical value of the energy gap at the Fermi level, and supplement results obtained in the previous section, the Eliashberg equations are solved in the mixed representation (for more details please see \cite{szczesniak2}, \cite{durajski2} and \cite{szczesniak9}). This procedure allows to obtain the solutions of the Eliashberg equations on the real axis ($\omega$), which provide the quantitative values of the energy gap at the Fermi level ($2\Delta(T)$) from:

\begin{equation}
\label{eq3}
\Delta\left(T\right)={\rm Re}\left[\Delta\left(\omega=\Delta\left(T\right),T\right)\right].
\end{equation}

\begin{figure}[ht]
\includegraphics[width=\columnwidth]{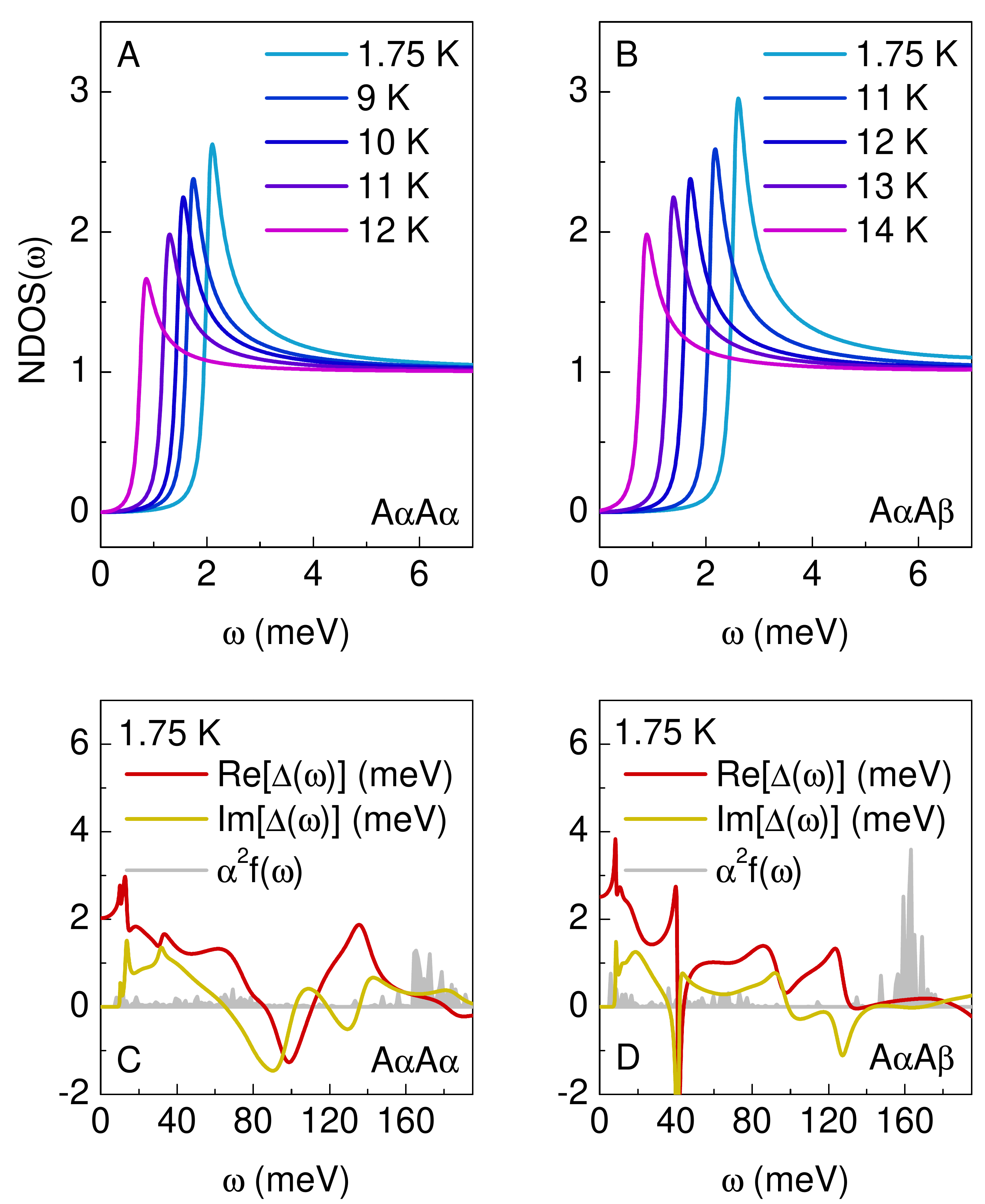}
\caption{The normalized density of states as a function of frequency (NDOS($\omega$)) for selected temperature values and two considered $A \alpha A \alpha$ (A) and $A \alpha A \beta$ (B) stacking cases. The real (Re$[\Delta(\omega)]$) and the imaginary (Im$[\Delta(\omega)]$) part of the order parameter on the real axis at $T=T_{0}=1.75$ K for the $A \alpha A \alpha$ (C) and $A \alpha A \beta$ (D) configurations. For comparison purposes, the corresponding Eliashberg functions are plotted in subfigures (C) and (D).}
\label{fig3}
\end{figure}

The obtained results are summarized in Fig \ref{fig3} (A) and (B) in the form of the total normalized density of states (NDOS$(\omega)$) for the selected values of temperature, calculated as:

\begin{equation}
\label{eq4}
{\rm NDOS}\left(\omega \right)=\frac{\rm DOS_{S}\left(\omega \right)}{\rm DOS_{N}\left(\omega \right)}={\rm Re}\left[\frac{\left|\omega -i\Gamma \right|}{\sqrt{\left(\omega -i\Gamma\right)^{2}}-\Delta^{2}\left(\omega\right)}\right],
\end{equation}
where $\rm DOS_{S}\left(\omega \right)$ and $\rm DOS_{N}\left(\omega \right)$ denote the density of states in the superconducting and normal state, respectively. Moreover, $\Gamma$ is the pair breaking parameter equal to $0.15$ meV.

In Fig \ref{fig3} (A) and (B) the thermal effect of the successive decrease of the $2\Delta(T)$ value with the increasing temperature can be observed. In the end the gapless behavior is reached at $T=T_{C}$, when the considered material reveals the metallic character.

The detailed data of the real and imaginary parts of the order parameter function at $T_{0}=1.75$ K are presented in Fig \ref{fig3} (C) and (D). These results suggest lack of the damping effects at the low-frequencies due to the non-zero values of the Re$[\Delta(\omega)]$ parts. Also, strong correlation between the shapes of the Re$[\Delta(\omega)]$, Im$[\Delta(\omega)]$ parts and the corresponding Eliashberg functions is clearly visible. 

The calculated values of the zero temperature energy gap at the Fermi level ($2\Delta(0)\equiv2\Delta(T_{0})$) are equal to $4.07$ meV and $5.09$ meV for the $A \alpha A \alpha$ and $A \alpha A \beta$ systems, respectively. Again, the notable enhancement can be observed in the reference to the corresponding energy gap value for the monolayer LiC$_{6}$. The appropriate ratios which describe this fact are: ${2\Delta(0)^{A \alpha A \alpha}}/{2\Delta(0)^{A \alpha}}=1.49$ and ${2\Delta(0)^{A \alpha A \beta}}/{2\Delta(0)^{A \alpha}}=1.86$, where $2\Delta(0)^{A \alpha}=2.74$ meV \cite{szczesniak2}. These results are close to the corresponding ones presented in the previous section for the superconducting transition temperature, confirming simultaneously the strong correlation between the transition temperature and the value of the energy gap at the Fermi level. Following the BCS theory, these two parameters can be represented in the form of the characteristic dimensionless ratio as \cite{bardeen1}, \cite{bardeen2}: $R_{\Delta}=2\Delta(0)/k_{B} T_{C}$. The $R_{\Delta}$ ratio is constant within the BCS limit and equals 3.53, whereas for both considered cases the obtained values are equal to 3.74 and 4.02 for the $A \alpha A \alpha$ and $A \alpha A \beta$ systems, respectively.

\begin{figure}[ht]
\includegraphics[width=\columnwidth]{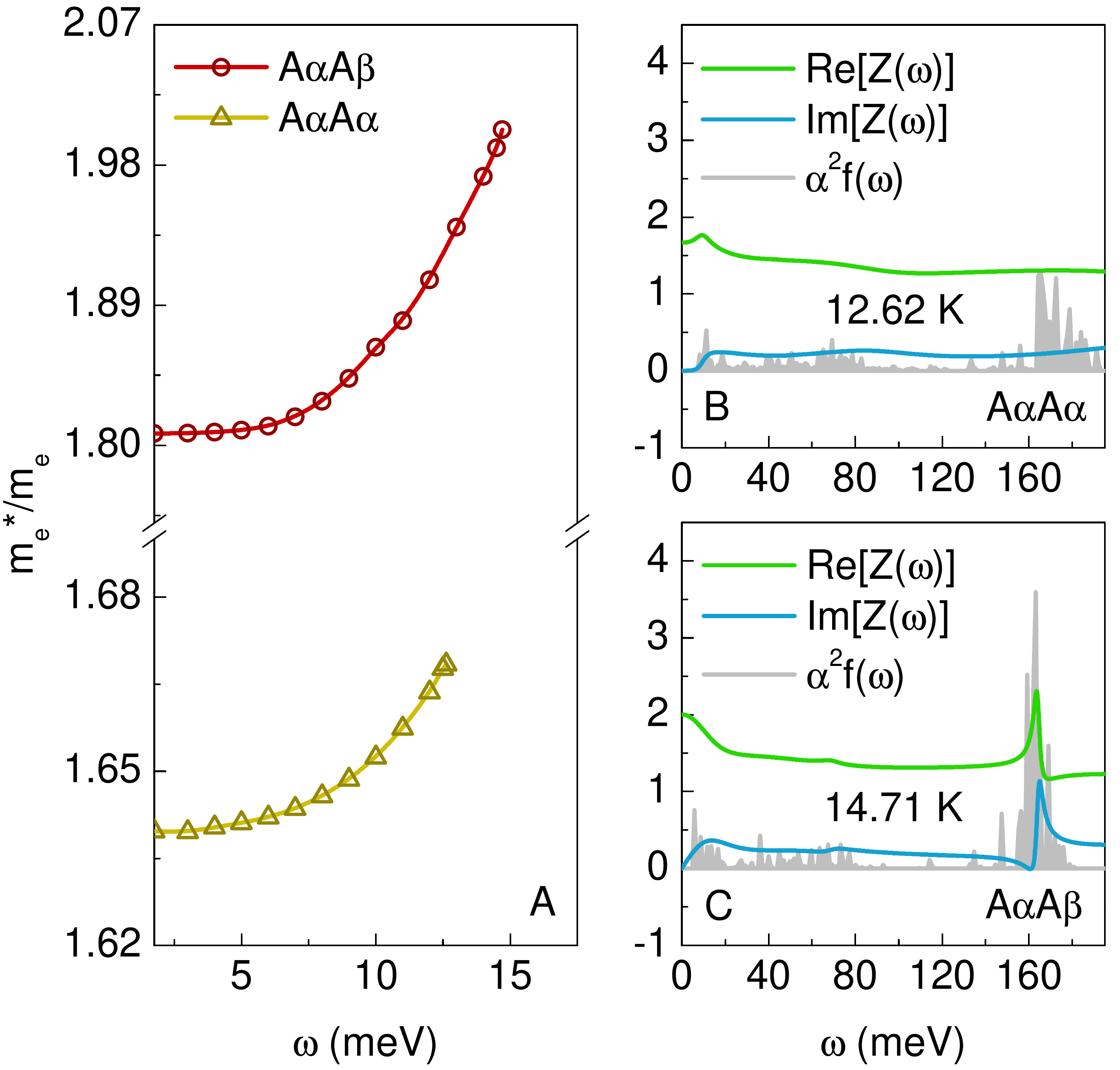}
\caption{(A) The effective electron mass to the bare electron mass ratio ($m_{e}^{*}/m_{e}$) as a function of temperature for the $A \alpha A \alpha$ and $A \alpha A \beta$ systems. The real (Re$[Z(\omega)]$) and the imaginary (Im$[Z(\omega)]$) part of the wave function renormalization factor at $T=T_{C}$ for the $A \alpha A \alpha$ (B) and $A \alpha A \beta$ (C) stacking cases. For comparison purposes, the corresponding Eliashberg functions are plotted in subfigures (B) and (C).}
\label{fig4}
\end{figure}

The exact calculation of the electron effective mass, follows the imaginary axis case, and based on the analysis of the wave function renormalization factor on the real axis ($Z(\omega)$), by using the following relation:

\begin{equation}
\label{eq5}
m_{e}^{*}=Re[Z(\omega=0)]m_{e},
\end{equation}

The $m_{e}^{*}/m_{e}$ ratios as a function of temperature are presented in Fig. \ref{fig4} (A) for two considered stacking cases of the lithium-doped bilayer graphene. Additionally, the corresponding detailed behavior of the Re$[Z(\omega)]$ function at $T=T_{C}$ for the $A \alpha A \alpha$ and $A \alpha A \beta$ systems is depicted in Fig. \ref{fig4} (B) and (C), respectively. The maximum value of the electron effective mass is obtained for the $A \alpha A \beta$ configuration at $T=T_{C}$ and equals $2m_{e}$, comparing to the $1.61m_{e}$ obtained for the LiC$_6$ monolayer in \cite{szczesniak2}. The comparison ratios between the LiC$_6$ monolayer and both considered in this paper stacking cases are: ${[m_{e}^{*}]^{A \alpha A \alpha}}/{[m_{e}^{*}]^{A \alpha}}=1.04$ and ${[m_{e}^{*}]^{A \alpha A \beta}}/{[m_{e}^{*}]^{A \alpha}}=1.24$, where $[m_{e}^{*}]^{A \alpha A \alpha}=1.67m_{e}$.

%%%%%%%%%%%%%%%%%%%%%%%%%%%%%%%%%%%%%%%%%%%%%%%%%%%%%%%
\subsection{Free energy and entropy difference between normal and superconducting state}
%%%%%%%%%%%%%%%%%%%%%%%%%%%%%%%%%%%%%%%%%%%%%%%%%%%%%%%

In the next step the free energy difference between the normal and superconducting state is calculated on the basis of the expression:
\begin{eqnarray}
\label{eq6}
\frac{\Delta F}{\rho\left(0\right)}&=&-\frac{2\pi}{\beta}\sum_{n=1}^{M}
\left(\sqrt{\omega^{2}_{n}+\Delta^{2}_{n}}- \left|\omega_{n}\right|\right)\\ \nonumber
&\times&(Z^{S}_{n}-Z^{N}_{n}\frac{\left|\omega_{n}\right|}
{\sqrt{\omega^{2}_{n}+\Delta^{2}_{n}}}),
\end{eqnarray}  
where the $\rho(0)$ denotes the electron density of states at the Fermi level, and $Z^{S}_{n}$ and $Z^{N}_{n}$ are the the wave function renormalization factors for the superconducting ($S$) and normal ($N$) state, respectively. The determination of the $\Delta F$ function is of crucial importance for the further calculations of the entropy difference between the superconducting and normal state ($\Delta S$), the specific heat for the superconducting state ($C_{S}$), and the thermodynamic critical field ($H_{C}$).

\begin{figure}[ht]
\includegraphics[width=\columnwidth]{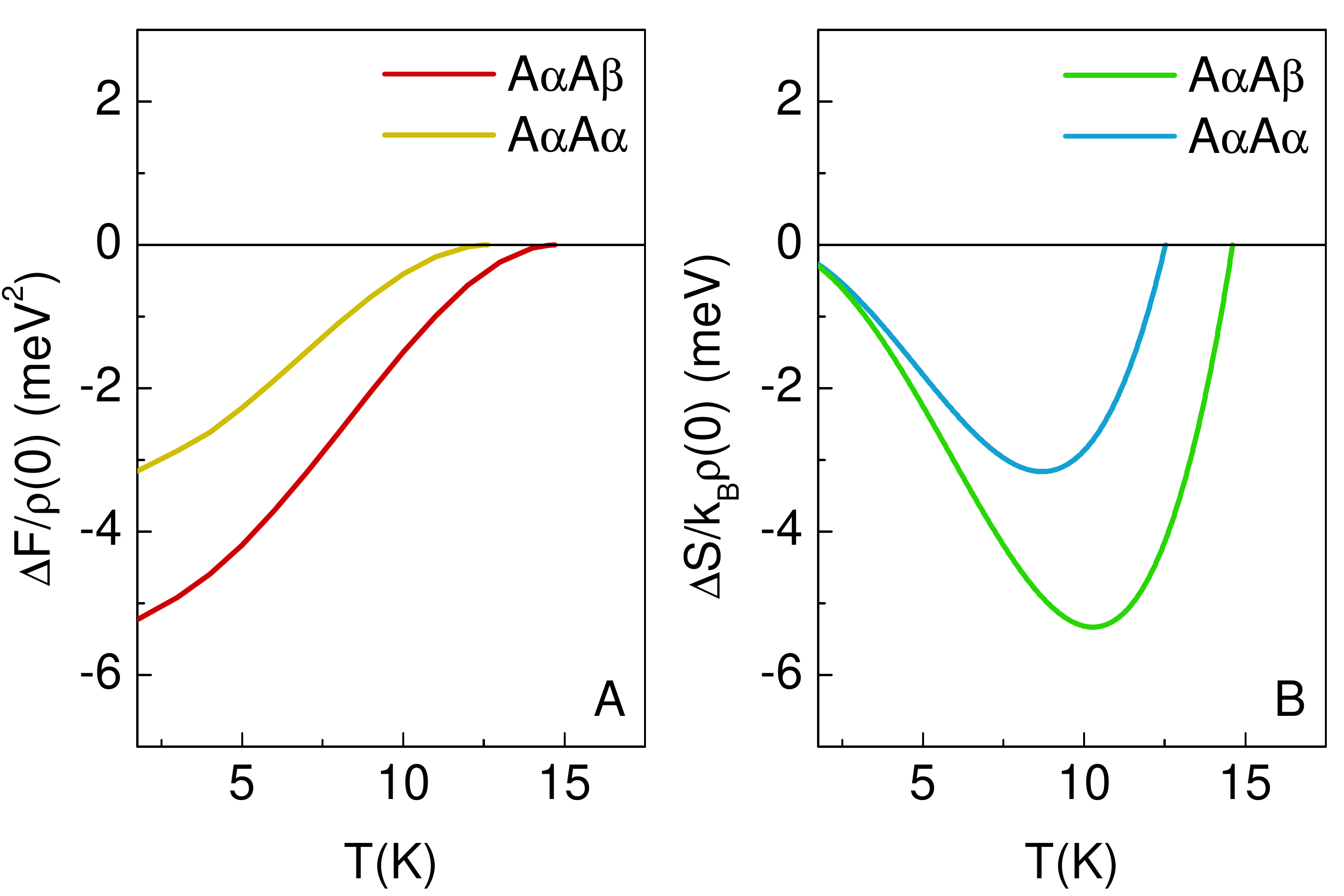}
\caption{(A) The normalized free energy difference between the superconducting and normal state ($\Delta F/\rho\left(0\right)$) as a function of temperature for the $A \alpha A \alpha$ and $A \alpha A \beta$ stacking cases. (B) The temperature dependance of the normalized entropy difference between the superconducting and normal state ($\Delta S/k_{B}\rho\left(0\right)$) for the $A \alpha A \alpha$ and $A \alpha A \beta$ configurations.}
\label{fig5}
\end{figure}

Aforementioned $\Delta F$ and $\Delta S$ differences between the superconducting and normal state are presented in Fig. \ref{fig5} as a function of temperature. The normalized $\Delta S/\rho(0)$ function is calculated as:

\begin{equation}
\label{eq7}
\frac{\Delta S}{k_{B}\rho\left(0\right)}=\frac{d\left[\Delta F/\rho\left(0\right)\right]}{d\left(k_{B}T\right)}.
\end{equation}

It can be seen that both functions take the negative values and reach value of zero at $T=T_{C}$. The negativity of the free energy difference between the normal and superconducting states, assures the thermodynamic stability of the superconducting state in the temperature range $T \in \left< T_{0}, T_{C} \right>$. In what follows, the change from the $A \alpha A \alpha$ to the $A \alpha A \beta$ stacking strengthens the the stability of the supercodnucting phase in the lithium-decorated bilayer graphene. At the same time, for both stacking cases, the $\Delta S(T_{C})/\rho(0)=0$ realtion satisfies the third law of thermodynamics. In summary, results presented in Fig. \ref{fig5} are the testimony for the correctness of the calculations at this point.

%%%%%%%%%%%%%%%%%%%%%%%%%%%%%%%%%%%%%%%%%%%%%%%%%%%%%%%
\subsection{Specific heat and thermodynamic critical field}
%%%%%%%%%%%%%%%%%%%%%%%%%%%%%%%%%%%%%%%%%%%%%%%%%%%%%%%

Solutions obtained in the previous section are further used to determine the temperature dependance of the specific heat for the superconducting state ($C_{S}$), and the thermodynamic critical field ($H_{C}$). In particular, the latter quantity is calculated as:
\begin{equation}
\label{eq8}
\frac{H_{C}}{\sqrt{\rho\left(0\right)}}=\sqrt{-8\pi\left[\Delta F/\rho\left(0\right)\right]}.
\end{equation}
On the other hand, the specific heat for the superconducting state is given by:
\begin{equation}
\label{eq9}
C^{S}=C^{N}+\Delta C.
\end{equation}
where, $C^{N}$ denotes the specific heat for the normal state of the given form:
\begin{equation}
\label{eq10}
\frac{C^{N}\left(T\right)}{ k_{B}\rho\left(0\right)}=\frac{\gamma}{\beta}, 
\end{equation}
and $\Delta C$ is the difference between the specific heat for the superconducting and normal state written as:
\begin{equation}
\label{eq11}
\frac{\Delta C\left(T\right)}{k_{B}\rho\left(0\right)}=-\frac{1}{\beta}\frac{d^{2}\left[\Delta F/\rho\left(0\right)\right]}{d\left(k_{B}T\right)^{2}}.
\end{equation}

\noindent In Eq. \ref{eq10}, $\gamma\equiv\frac{2}{3}\pi^{2}\left(1+\lambda\right)$ is the Sommerfeld constant.

\begin{figure}[ht!]
\includegraphics[width=\columnwidth]{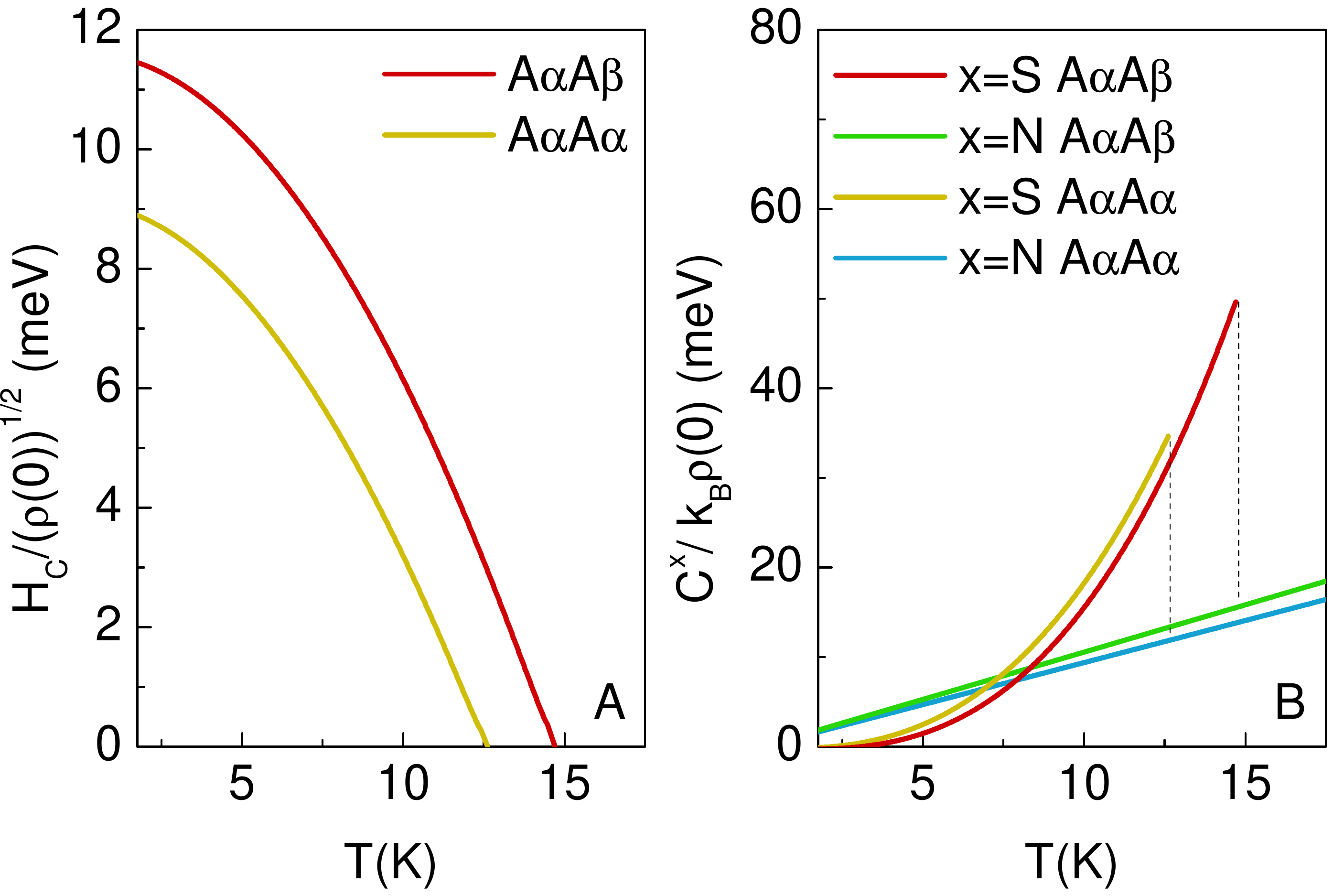}
\caption{(A) The temperature dependance of the normalized thermodynamic critical field ($H_{C}/\sqrt{\rho\left(0\right)}$) for the $A \alpha A \alpha$ and $A \alpha A \beta$ stacking cases. (B) The normalized specific heat for the superconducting ($C^{S}/k_{B}\rho\left(0\right)$) and normal ($C^{N}/k_{B}\rho\left(0\right)$) state as a function of temperature for two considered $A \alpha A \alpha$ and $A \alpha A \beta$ configurations.}
\label{fig6}
\end{figure}

The results obtained for the temperature dependance of the normalized thermodynamic critical field ($H_{C}/\sqrt{\rho\left(0\right)}$) and the normalized superconducting specific heat ($C^{S}/k_{B}\rho\left(0\right)$) are presented in Fig. \ref{fig6} (A) and (B) respectively. Particularly, in Fig. \ref{fig6} (B), the characteristic {\it jump} of the $C^{S}/k_{B}\rho\left(0\right)$ function at $T=T_{C}$ is visible and marked by the vertical dashed lines.

Futhermore, the maximum values for the normalized thermodynamic critical field ($H_{C}(0)$) amount 9.07 meV and 11.60 meV, whereas for the normalized specific heat for the superconducting state ($C^{S}(T_{C})$) equal 34.67 meV and 49.65 meV for the $A \alpha A \alpha$ and $A \alpha A \beta$ systems, respectively. In comparison to the LiC$_{6}$ monolayer case the obtained ratios are: ${H_{C}(0)^{A \alpha A \alpha}}/{H_{C}(0)^{A \alpha}}=1.45$ and ${H_{C}(0)^{A \alpha A \beta}}/{H_{C}(0)^{A \alpha}}=1.86$, where $H_{C}(0)^{A \alpha}=6.24$ meV, and ${C^{S}(T_{C})^{A \alpha A \alpha}}/{C^{S}(T_{C}))^{A \alpha}}=1.80$ and ${C^{S}(T_{C})^{A \alpha A \beta}}/{C^{S}(T_{C})^{A \alpha}}=2.58$, where $C^{S}(T_{C})^{A \alpha}=19.28 $ meV.

Above results allows next the determination of the characteristic dimensionless ratios for the thermodynamic critical field ($R_{H}$) and specific heat ($R_{C}$). In particular, the $R_{H}$ and $R_{C}$ parameters are given by \cite{bardeen1}, \cite{bardeen2}:
\begin{equation}
\label{r14}
R_{H}\equiv\frac{T_{C}C^{N}\left(T_{C}\right)}{H_{C}^{2}\left(0\right)},
\quad {\rm and} \quad
R_{C}\equiv\frac{\Delta C\left(T_{C}\right)}{C^{N}\left(T_{C}\right)}.
\end{equation}
In what follows, the obtained estimations are: R$_{\rm H}$=0.163 and R$_{\rm C}$=1.93 for $A \alpha A \alpha$, and R$_{\rm H}$=0.150 and R$_{\rm C}$=2.20 for $A \alpha A \beta$, where the BCS theory predicts: R$_{\rm H}$=0.168 and R$_{\rm C}$=1.43.

%%%%%%%%%%%%%%%%%%%%%%%%%%%%%%%%%%%%%%%%%%%%%%%%%%%
\section{Summary}
%%%%%%%%%%%%%%%%%%%%%%%%%%%%%%%%%%%%%%%%%%%%%%%%%%%
 
The present paper reports systematic analysis of the superconducting phase in the lithium-decorated bilayer graphene with two different layer stacking configurations ($A \alpha A \alpha$ and $A \alpha A \beta$). The presented discussion highlights quantitatively subtle relation between the composition and arrangement of the lithium adatoms on the graphene sheets and the resulting superconducting properties. In general, the superconducting graphene-lithium nanosystems are then expected to be rather sensitive to the material engineering and further preparation of the corresponding nanodevices.

The main finding of the present paper reveals that the analyzed lithium-decorated bilayer graphene systems exhibit notable enhancement of the superconducting thermodynamic properties in comparison to the monolayer case discussed previously in \cite{profeta} and \cite{szczesniak2}. Of particular attention is the $A \alpha A \beta$ stacking configuration case which exhibit stronger superconducting properties then the $A \alpha A \alpha$ one and is characterized by the transition temperature equals to 14.71 K.

It is also proved that the $A \alpha A \alpha$ and $A \alpha A \beta$ nanosystems can be analyzed only within the strong-coupling regime. This observation is made on the basis of the comparison between the characteristic thermodynamic ratios ($R_{\Delta}$, $R_{H}$, and $R_{C}$) calculated for the considered systems and the values predicted by the BCS theory. In particular, the $R_{\Delta}$, $R_{H}$, and $R_{C}$ exceeds limits of the BCS theory for both the $A \alpha A \alpha$ and $A \alpha A \beta$ stacking cases. These discrepancies occur due to the strong-coupling and retardation effects, which are not included in the mean-field BCS theory. In what follows, the BCS approach is likely inadequate for the proper analysis of the superconducting properties in the context of the bilayer lithium-graphene systems.

%%%%%%%%%%%%%%%%%%%%%%%%%%%%%%%%%%%%%%%%%%%%%%%%%%%
\begin{acknowledgments}

Author is thankful to David M. Guzman (Purdue University), Hamad M. Alyahyaei (University of California) and Radi A. Jishi (California State University) for sending him data prior to publication, and to Rados{\l}aw Szcz{\c{e}}{\'s}niak (Cz{\c{e}}stochowa University of Technology and Jan D{\l}ugosz University) for fruitful scientific discussions throughout the work on the present paper.

\end{acknowledgments}
%%%%%%%%%%%%%%%%%%%%%%%%%%%%%%%%%%%%%%%%%%%%%%%%%%%
\bibliographystyle{apsrev}
\bibliography{manuscript}
%%%%%%%%%%%%%%%%%%%%%%%%%%%%%%%%%%%%%%%%%%%%%%%%%%%
\end{document}